\documentclass{sig-alternate}
\usepackage{graphicx}
\usepackage{wrapfig}
\usepackage[all]{xy}

\long\def\comment#1{}
\newcommand{\commentout}[1]{}
\newcommand{\denselist}{
      \setlength{\itemsep}{0pt}
      \setlength{\parsep}{1.5pt}
      \setlength{\topsep}{1.5pt}
      \setlength{\parskip}{2pt}
      \setlength{\partopsep}{0pt}
      \setlength{\labelwidth}{1em}
      \setlength{\labelsep}{0.5em} }
\newcommand{\bdesc}{\begin{description}\denselist}
\newcommand{\edesc}{\end{description}}

\newcommand{\ie}{{\em i.e.}}

\newcommand{\secref}[1]{Section~\ref{#1}}
\newcommand{\figref}[1]{Figure~\ref{#1}}

\begin{document}
\conferenceinfo{The 2nd SNA-KDD Workshop '08 ( SNA-KDD'08)}{August 24, 2008 , Las Vegas, Nevada,USA.)}
\CopyrightYear{2008} 
\crdata{978-1-59593-848-0}  

\title{Community Detection using a Measure of Global Influence}
\numberofauthors{2}
\author{
\alignauthor
Rumi Ghosh\\
       \affaddr{University of Southern California}\\
       \affaddr{Information Sciences Institute}\\
       \affaddr{4676 Admiralty Way}\\
      \affaddr{Marina del Rey, California 90292}\\
       \email{rumig@usc.edu}
\alignauthor
Kristina Lerman\\
      \affaddr{University of Southern California}\\
       \affaddr{Information Sciences Institute}\\
       \affaddr{4676 Admiralty Way}\\
       \affaddr{Marina del Rey, California 90292}\\
       \email{lerman@isi.edu }
}

\maketitle 

\begin{abstract}
The growing popularity of online social networks has provided researchers with access to large amount of social network data. This, coupled  with the  ever increasing computation speed, storage capacity and data mining capabilities, led to the renewal of interest in automatic community detection methods. Surprisingly, there is no universally accepted definition of the community. 
One frequently used definition states that ``communities, that have more and/or better-connected `internal edges' connecting members of the set than `cut edges' connecting the set to the rest of the world''~\cite{Leskovec08www}.
This definition inspired the modularity-maximization class of community detection algorithms, which look for regions of the network that have higher than expected density of edges within them.
We introduce an alternative definition which states that a community is composed of individuals who have more influence on others within the community than on those outside of it.
We present a mathematical formulation of influence, define an influence-based modularity metric, and show how to use it to partition the network into communities. We evaluated our approach on the standard data sets used in literature, and found that it often outperforms the edge-based modularity algorithm.
\end{abstract}


\section{Introduction} \label{sec:intro}
\comment{
A man is judged by the company he keeps. Our friends and the social communities to which we belong have a deep influence on our lives and go a long way to define who we are and what we do. Who we are and what we do in turn influence our choice of the communities we join, and how we influence those communities.
The desire to be recognized by, and exert influence, over other community members drives us to seek contact with the influential people of the  community who could help us increase our overall influence within the community. However, merely having a lot of contacts does not imply a greater degree of influence. The degree of influence exerted by an individual on the community depends not only on \emph{how many} contacts she has, but also \emph{who} the contacts are~\cite{Katz,PageRank}.
}
Communities and social networks have been a source of interest for researchers for several decades ~\cite{Pool,Katz}. However, one of the main problems faced by the early researchers was the difficulty of collecting acquaintanceship and related empirical data from human subjects~\cite{Pool}. The advent of the internet and the growing popularity of online social networks changed that, providing the researchers access  to huge amount of invaluable human social network data. This, coupled  with the  ever increasing computation speed, storage capacity and data mining capabilities, led to the reemergence of interest in the social networks in general, and community detection methods specifically.

Despite a long history of investigation, surprisingly, there is not a single universally accepted definition of the community. A  definition preferred by sociologists is that a community is composed of individuals who are similar to one another in some way, whether it is because they see the same friends or belong to the same organizations. This definition inspired the class of community-finding methods based on hierarchical clustering. These algorithms assign nodes to the same community if they are sufficiently similar to each other. Similarity measures include structural equivalence, where two nodes are said to be equivalent if they have the same set of neighbors, and approximate equivalence that uses Euclidean distance and Pearson correlation. Another similarity measure used in hierarchical clustering methods is the number of paths between nodes. Hierarchical clustering, however, may not assign every node to a non-trivial community. In addition, it does not provide a measure of how good a particular division of the network into communities is.

Physicists and computer scientists prefer to define community as ``a group of vertices in which there are more edges between vertices within the group than to vertices outside of it''~\cite{Clauset}. This definition helped inform a variety of graph-based approaches to automatic community detection, including graph partitioning and modularity optimization techniques.  Graph partitioning algorithms~\cite{Fiedler,Pothen} attempt to minimize the number of edges running between communities.
One of the main disadvantages of these methods is that either the  number of communities has to be specified \emph{ a priori}, or they repeatedly bisect the graph without a well-defined stopping point. Since it is almost impossible to always know beforehand the number of communities within a large network, these methods are unable to automatically detect natural communities. Furthermore there is no guarantee that the communities into which we have divided the network represent the best possible community division of the network.
Newman and his colleagues realized that rather than minimize the number of edges running between groups, one should instead look for groups that have higher than expected number of edges within them and lower than expected edges between them~\cite{Newman104,Newman204,Newman106,Newman206}. These algorithms maximize a measure called \emph{modularity}, which is the fraction of all edges within communities minus the expected value of the same quantity.
The modularity optimization method is fast (if approximate), and can be applied to both undirected and directed graphs. It is able to find the ``best'' assignment of nodes to communities, although each node can belong  to only a single community.
Some researchers have recently questioned the applicability to real-world networks of the edge-density definition of the community and the modularity optimization techniques based on it. Leskovec et al.~\cite{Leskovec08www} found that in large networks, communities tend to `blend' into the giant connected component, making it impossible to extract any but the trivial small and tightly knit communities.

We stake a claim in this active field by introducing an alternative definition of community that is based on information spread on networks. We claim (without much theoretical or empirical support) that a community is composed of individuals who have more influence on individuals within the community than on those outside of it. We take a structure-based view of influence, defining it as the number of paths, of any length, that exist between two nodes. The more paths there are, the more opportunities one node has to affect the other. This will result in the actions of the community members becoming correlated with time, whether through adopting a new fashion trend or vocabulary terms, watching a movie, or buying a product. We define influence-based modularity metric, and show how to use it to partition a network into communities. We evaluated our approach on the standard data sets used in literature, and found that it gives at least as good performance as the standard modularity-based algorithm. 

The paper is organized as follows. In \secref{sec:influence} we define and give a mathematical derivation of influence. \secref{sec:modularity} describes our re-definition of the modularity metric in terms of influence, and shows how the new modularity can be used for automatic community detection. We present results of applying our approach to well-studied networks in \secref{sec:evaluation}. In \secref{sec:related} we compare our approach to those that have previously been described in literature, and conclude with \secref{sec:conclusion}.

\section{A Measure of  Global Influence}
\label{sec:influence}
A network  of $N$ nodes and  $E$ links can be represented  using a graph $G(N,E)$, where  $N$ is the number of vertices of the graph representing the nodes of the network, and $E$ is the number of edges of the graph. Edges are directed; however, if there exists a an edge from vertex $i$ to $j$ and also from $j$ to $i$, it is represented as an undirected edge. A path $p$  is an \emph{n  hop path} from  vertex $i$  to $j$,  if there are $n$ vertices between the vertex $i$ and vertex $j$  along the path. We allow the paths to be non-selfavoiding, meaning that the same pair of vertices could be traversed more than once on the path. The graph $G(N,E)$ can be represented by an adjacency matrix $\mathbf{A}$ whose elements $A_{ij}$ are defined as
 \[
A_{ij} = \left \{ \begin {array} {ll}
1 & \mbox{ if  $ \exists$  an edge from vertex $i$ to $j$  } ; \\
0 & \mbox{ otherwise}.
\end{array}
\right. \]

We introduce an index for measuring the degree of \emph{influence} a node has on other nodes. We use this index to divide the network into communities  so that nodes which have higher influence on each other are grouped together. At the same time this index could also be used to find out the status of the people in the community based on their influence~\cite{Katz}.

  \begin{figure}[tbh]
 \includegraphics[height= 2.1 in]{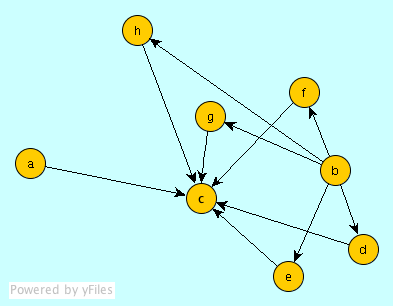}
 \caption{Connectivity:Edge Connectivity and Path Connectivity}
 \label{fig:conn}
 \end{figure}

Influence can be defined as the capacity to have an effect on someone. Pool and Kochen~\cite{Pool}  state that  ``influence in large part is the ability to reach a crucial man through the right channels, and the more the channels in reserve the better.'' This is the measure of global influence that we employ, and we also adopt the concept of attenuation when transmitted  through intermediaries~\cite{Katz}. Therefore, influence depends not only on direct contact between people, but also on the number of ways an individual can reach another, or the number of \emph{n hop} paths between them. Hence, the influence of node $a$ on $b$ is likely to be more if there are more paths from $a$ to $b$.

The strength of the effect  via longer paths with more intermediaries is likely to be lower than via shorter chains with fewer intermediaries. We model the attenuation of influence over longer chains through two parameters $ \alpha $   and $\beta$. We use two parameters, rather than a single parameter, to model the fact that a node may have more influence over its direct neighbors, than it will have over the neighbor's neighbors, and so on. Thus, $\beta $ ($ 0 \le \beta \le 1$) is the \emph{direct attenuation factor},  the probability that the effect will be transmitted to the immediate neighbors of the node.  $\alpha$ ($ 0 \le \alpha \le 1$) is the \emph{indirect attenuation factor}, the probability that the effect will be transmitted through links other than those to the node's immediate neighbors (i.e., via friends of friends). Let us consider transmitting an effect or a message from node $b$  to node $c$ in a network in \figref{fig:conn}. The probability of transmission to the immediate neighbors of $b$ is $\beta$. The probability of transmission over the five $1$-hop paths is $\beta \alpha$. In general, the probability of a transmission along an $n$-hop path is $\beta \alpha^{n-1}$. Note that $\beta =\alpha$ is a special case when the transmission probability along all links is the same.

The total influence of node $b$ on node $c$ is thus dependent on the number of (attenuated) channels between $b$ and $c$, or  the sum of all the  weighted paths from  node $b$ to $c$. This definition of influence makes intuitive sense, because the greater the number of paths between $b$ and $c$, the more opportunities there are for $b$ to transmit messages to $c$ and to affect what $c$ is doing.

We represent total number of links from node $i$ to node $j$ as \xymatrix{ i\ar[r]^{0} & j}, which is given by the elements $ A_{ij}$ of the adjacency matrix $\mathbf{A}$.
Next, we represent the total number of $1$-hop paths from node $i$ to node $j$ as  \xymatrix{ i\ar[r]^{1} & j}, and it is given by $\sum_{k=1}^{N} A_{ik} A_{kj} $, since a path can exist from $i$ to $j$  in one hop via a particular node $k$ iff $\exists$ an edge from  $i$ to $k$ and also from $k$ to $j$. Summing over all $k \in N$, we get the above result. We define matrix $\mathbf{A_1 =A \cdot A}$ whose elements, ${A_1}_{ij}= \sum_{k=1}^{N} A_{ik} A_{kj}$, give the the total number $1$-hop paths from $i$ to $j$.

Similarly  the total number of $2$-hop paths from node $i$ to node $j$ is represented  as \xymatrix{ i\ar[r]^{2} & j}  and is given by \begin{equation*}
\sum_{l=1}^{N} {(\sum_{k=1}^{N} A_{ik} A_{kl})} A_{lj}\,. 
\end{equation*}
We define matrix $\mathbf{A_2=A \cdot A \cdot A}$ whose elements
${A_2}_{ij} = \sum_{l=1}^{N} {(\sum_{k=1}^{N} A_{ik} \cdot A_{kl})} \cdot A_{lj}$ give the the total number $2$-hop paths from $i$ to $j$.

Generalizing  the total number of chains from node $i$ to node $j$ with $n$ intermediaries  $j$  is represented  as \xymatrix{ i\ar[r]^{n} & j} and  is  by the matrix $\mathbf{A_n}$ where
\begin{equation}
\mathbf{A_n =\overbrace{A\cdot A\cdots A}^{n+1\ times}=A_{(n-1)}
 \cdot A}
\end{equation}

Adding weights to take into account the attenuation of effect of node $i$ on node $j$, we get total influence of node $i$ on $j$ as
\begin{eqnarray*}
\xymatrix{ i\ar[r]& j} = \beta \xymatrix{ i\ar[r]^{0} & j} & + & \beta \alpha \xymatrix{ i\ar[r]^{1} & j}+\cdots  \\
& + & \beta \alpha^{n}\xymatrix{ i\ar[r]^{n} & j} \cdots
\end{eqnarray*}

We  represent  the measure of influence of nodes on other nodes by the \emph {influence matrix} $\mathbf {P}$ where
 \begin{equation}
 \label{eq:inf}
 \mathbf{P= \beta A +\beta \alpha A_1 + \cdots + \beta \alpha^n  A_n+\cdots }\,.
\end{equation}
\comment{
Multiplying both sides of Equation\eqref{eq:inf}  by $\alpha A$
 we get
 \begin{equation}
 \label{eq:inf2}
 \mathbf {P\alpha A = \beta\alpha A.A +\beta \alpha^2 A.A.A+ \cdots + \beta \alpha^{n+1}  \overbrace{A. A\cdots.A}^{n+1 times}+\cdots}
 \end{equation}
 Now substracting Equation ~\eqref{eq:inf2} from Equation\eqref{eq:inf1} we get
 \begin{equation}
\label {eq:inf3}
\mathbf{P( I-\alpha A) = \beta A}
\end{equation}
which implies
}
After elementary manipulations, this series can be rewritten as
\begin{equation}
\label {eq:inf4}
\mathbf{P = \beta A {( I-\alpha A)}^{-1}}
\end{equation}
where $I$ is the identity matrix. This equation holds while  $\alpha$ is less than the reciprocal of largest characteristic root of adjacency matrix  $\mathbf{A}$ \cite{Ferrar}.

The influence matrix captures the effective connectedness of a node not only in terms of the number of nodes  it  is directly connected to, but also in terms the number of nodes it is indirectly connected to.
This formulation is mathematically similar to the weights between vertices used in the hierarchical clustering  algorithm of Girvan and Newman~\cite{GirvanNewman02}, where the weights depended on the total number of paths between nodes.
Rather than using influence to measure similarity between nodes, as done in that work, we will use it to find groups of nodes that exert higher than expected influence on each other.


\section{Communities and Influence}
\label{sec:modularity}

The objective of the algorithms proposed by  Newman and coauthors was to discover ``community structure in networks --- natural divisions of network nodes into densely \emph{connected} subgroups''~\cite{GirvanNewman04}. They proposed \emph{modularity} as a measure for evaluating the strength of the discovered community structure. Algorithmically, their approach to discovering network structure is based on finding groups with higher than expected edges within them and lower than expected edges between them~\cite{Newman104,Newman204,Newman106,Newman206}. The modularity $Q$, which is optimized by the algorithm is given by:\\
 $Q=$(fraction of edges within community)-(expected fraction of such edges). \\
Thus, they use $Q$ as a numerical index to evaluate a particular division of the network. The underlying idea, therefore, is that connectivity of nodes belonging to the same community is greater than that of nodes belonging to different communities, and they take the number of edges as the measure of connectivity. But is edge connectivity the true measure of connectivity on the network?

Consider again the graph in \figref{fig:conn}, where there exists an edge between $a$ and $c$ but not between $b$ and $c$. However, clearly $c$ is not unconnected from $b$, as there exist several distinct channels for $b$ to send information to, or influence, $c$.
The influence matrix that we defined above, gives a mathematical model of the global connectivity of the network. We will use this connectivity to identify communities in the network.

\subsection{Influence-based Modularity}
We redefine modularity $Q$ that as \\
$Q=$(connectivity within the community) - ( expected connectivity within the community)\\
and adopt the influence matrix $\mathbf{P}$ as the measure of connectivity.
This definition of modularity implies that in the best division of the network, the influence of nodes within their community is more than their influence outside their community. A division of the network into communities, therefore, maximizes the difference between the actual influence and the expected  influence within the community, given by the influence in an equivalent random graph.
Let us denote the expected influence by a $N \times N$ matrix $\mathbf{\bar{P}}$.
Modularity $Q$ then can be expressed as
\begin{equation}
\label{eq: mod2}
Q=\sum_{ij} {[P_{ij} - \bar{P_{ij}}]\delta(s_i, s_j)}
\end{equation}
where $s_i$ is the index of the community $i$ belongs to and 
\[
\delta(s_i, s_j) = \left \{ \begin {array} {ll}
1 & \mbox{ $s_i =s_j$} ; \\
0 & \mbox{ otherwise}.
\end{array}
\right. \]
When all the vertices are placed in a single group, then it is axiomatically assumed that $Q=0$. Thus we have $ \sum_{ij}[P_{ij} - \bar{P_{ij}}] =0$. Hence, the total influence $W$ is
\begin{equation}
\label{eq: mod3}
 W = \sum_{ij} \bar{P_{ij}}=\sum_{ij} P_{ij}
\end{equation}
Hence the null model against which we compare our network has the same number of vertices $N$ as the original model, and in it  the expected influence of the entire network equals to the actual influence of the original network. 

We further restrict the choice of null model to that where the expected  influence $W_j^{in}$ on a given vertex $j$ from all other vertices  is equal to the actual influence on the corresponding vertex in the real network.
\begin{equation}
\label{eq: mod4}
W_{j}^{in} = \sum_{i} \bar{P_{ij}} = \sum_{i} P_{ij}
\end{equation}
Similarly, we also assume that in the null model, the expected  influence $W_{i}^{out}$ of a given vertex $i$ on all other vertices  is equal to the actual influence of the corresponding vertex in the real network
\begin{equation}
\label{eq: mod5}
W_{i}^{out} = \sum_{j} \bar{P_{ij}} = \sum_{j} P_{ij}
\end{equation}
The null model of this class that we then consider has paths that are placed at random between vertices subject to the constraints  of Equation\eqref{eq: mod4} and  Equation ~\eqref{eq: mod5}.
This implies then that the expected influence $\bar{P_{ij}}$ of vertex $i$ on vertex $j$ can be written as
\begin{equation}
\label{eq: mod11}
\bar{P_{ij}} = f_1(W_{i}^{out})f_2(W_{j}^{in})\,,
\end{equation}
where $f_1$ and $f_2$ are some functions. We rewrite Equation\eqref{eq: mod5}  as
\begin{equation}
\label{eq: mod6}
W_{i}^{out} = 
\sum_{j}{ f_1(W_{i}^{out})f_2(W_{j}^{in})}= f_1(W_{i}^{out})\sum_{j}{f_2(W_{j}^{in})}
\end{equation}
for all $i$, and hence
\begin{equation}
\label{eq: mod12}
 f_1(W_{i}^{out}) = C_{1}{W_{i}^{out}}
\end{equation}
for some constant $C_1$.

Along the same lines we have
\begin{equation}
\label{eq: mod7}
W_{j}^{in} = 
\sum_{i}{ f_1(W_{i}^{out})f_2(W_{j}^{in})}= f_2(W_{j}^{in})\sum_{i}{f_1(W_{i}^{out})}
\end{equation}
for all $j$, and hence
\begin{equation}
\label{eq: mod13}
 f_2(W_{j}^{in})= C_{2}{W_{j}^{in}}
\end{equation}
for some constant $C_2$. Therefore, expected influence is 
\begin{eqnarray*}
\sum_{ij} \bar{P_{ij}}& =&\sum_{ij} (C_{1}C_{2} {W_{i}^{out}}{W_{j}^{in}})  \\
& =&C_{1}C_{2} \sum_{ij} ({W_{i}^{out}}{W_{j}^{in}})  \\
& =&C_{1}C_{2} {(\sum_{ij} P_{ij})}^2  \\
& =& C_{1}C_{2} W^{2}
\end{eqnarray*}
Now using Equation\eqref{eq: mod3} we have
\begin{equation}
\label{eq: mod9}
 W=  \sum_{ij} P_{ij} =\sum_{ij} \bar{P_{ij}} = C_{1}C_{2} W^{2}\,,
\end{equation}
which we can solve for $C_1C_2$.
Using Equations~\ref{eq: mod11}--\ref{eq: mod13} we can write expected influence as
\begin{equation}
\label{eq: mod14}
\bar{P_{ij}} = \frac { {W_{i}^{out}} {W_{j}^{in}}}{W}\,,
\end{equation}
and the influence-based modularity as
 \begin{equation}
\label{eq: mod15}
 Q=\sum_{ij} {\big[P_{ij} - \frac { {W_{i}^{out}} {W_{j}^{in}}}{W}\big]\delta(s_i, s_j)}
 \end{equation}

\subsection{Detecting Community Structure}
Once we have derived $Q$, we have to select an algorithm to divide the network into communities that optimize $Q$. Like others~\cite{ Newman206,Newman106,Leicht}, we  use the matrix-based approach analogous to spectral  partitioning. The possible approaches that could be then used for community detection include leading eigenvector method, vector partitioning method and so on. We implemented the leading eigenvector method~\cite{Newman206}.
We summarize the approach in the Appendix.

\section{Evaluation on Real Networks}
\label{sec:evaluation}

We evaluated our approach by using it to find communities on real networks that can be found in literature.

\subsection{Zachary's Karate Club}

\begin{figure*}[tbh]
\begin{tabular}{cc}
  \includegraphics[width=0.5\textwidth]{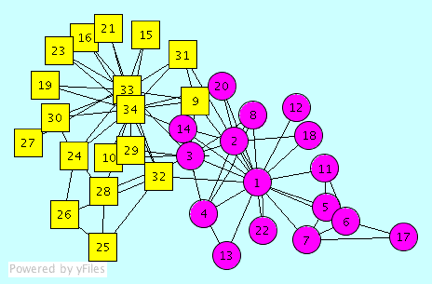} &
    \includegraphics[width=0.5\textwidth]{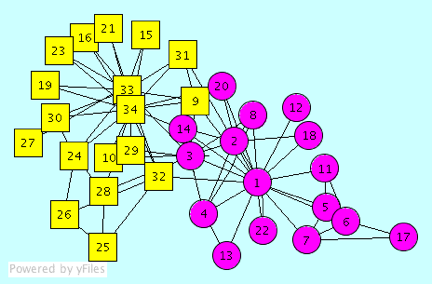}\\
  (a) & (b) \\
 \includegraphics[width=0.45\textwidth]{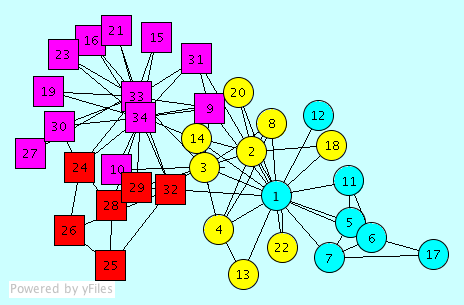} &
 \includegraphics[width=0.45\textwidth]{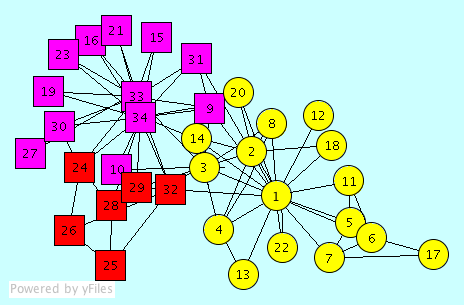}\\
(c) & (d) 
\end{tabular}
\caption{ Results of applying different community finding algorithms to Zachary's karate club network. The numbered vertices represent the members of the club and edges represent friendships. The factions in which the clubs split up during the course of study are shown by squares and circles. 
(a \& b) Communities found after running a single iteration (graph bisection) using (a) Newman's and (b) the proposed algorithms.
(c \& d) Natural communities found by running (c) Newman's algorithm and (d) the proposed algorith until termination condition is reached.
}
 \label{fig:1}
\end{figure*}

We applied this method on the friendship network of Zachary's karate club~\cite{Zachary}. In this study, Zachary studied the friendship network of a karate club for two years. During the course of the study, a disagreement developed between the administrator of the club and the club's instructor, resulting in the division of the club into two factions, represented by circles and  squares in \figref{fig:1}. The natural communities existing in the club has been predicted by various community detection and graph partitioning algorithms.
We used the friendship network of Zachary's karate club~\cite{Zachary} to compare the performance of the algorithm  proposed in this paper to Newman's community-finding algorithms. 

\figref{fig:1} presents results of different community-finding approaches. \figref{fig:1}(a) shows results of the  modularity maximization-based approach proposed by Newman~\cite{Newman204}  when the network is bisected into two communities only. \figref{fig:1}(b) shows results of a similar bisection done by our algorithm with $\beta= 1/N $ and $\alpha=1/N$, where $N=34$ is the number of nodes. Both methods result in the correct assignment of individuals to communities and are better than those produced by the spectral bisection algorithm and hierarchical clustering, which does not assign all nodes to the principal communities~\cite{Newman206}.
However, finding natural communities in the karate club network by iterating each algorithm until a stopping condition is reached, leads to different results. Newman's method divides the network into four communities (\figref{fig:1}(c)), while our method divides it into three communities (\figref{fig:1}(d)). Two of the communities generated by Newman's algorithm (shown in pink and red in \figref{fig:1}(c))  are similar to the two of the three communities found by our algorithm. However, it further subdivides the circle nodes into 
putting node $1$ into the same community as five of its immediate contacts, but a different community than nine of its immediate contacts.
Our algorithm appears to give a more realistic division of the karate club network into natural communities.

\subsection{College Football}

We also ran our approach on the US College football data from Girvan et al.~\cite{GirvanNewman02}\footnote{The college football data is available at \texttt{http://www-personal.umich.edu/$\sim$mejn/netdata/}.} The network represents the schedule of Division 1 games for the 2000 season where the vertices represent teams (colleges) and the edges represent the regular season game between the two teams they connect. The teams are divided into ``conferences'' containing 8 to 12 teams each. Games are more frequent between members of the same conference than members of different conferences leading to a  community structure with greater connectivity within the communities (represented by conferences) than between them. Inter-conference games however are not uniformly distributed, with teams that are geographically closer likely to play more games with one another than teams separated by geographic distances. However the as the authors state~\cite{GirvanNewman02} there are some conferences  like Sunbelt having teams playing nearly as many games against teams in other conferences (Western Athletic in case of Sunbelt) as they did against teams within their own conference. This leads to the intuition, that the conferences then may not be the natural communities present in given data, but the natural communities may actually be bigger than the the size of the conferences, with conferences playing as many games within them as between them being clubbed into the same community. How then can evaluate the purity of the natural communities detected?

 \begin{figure}[tbhp]
 \begin{tabular}{c}
   \includegraphics[width=0.55\textwidth]{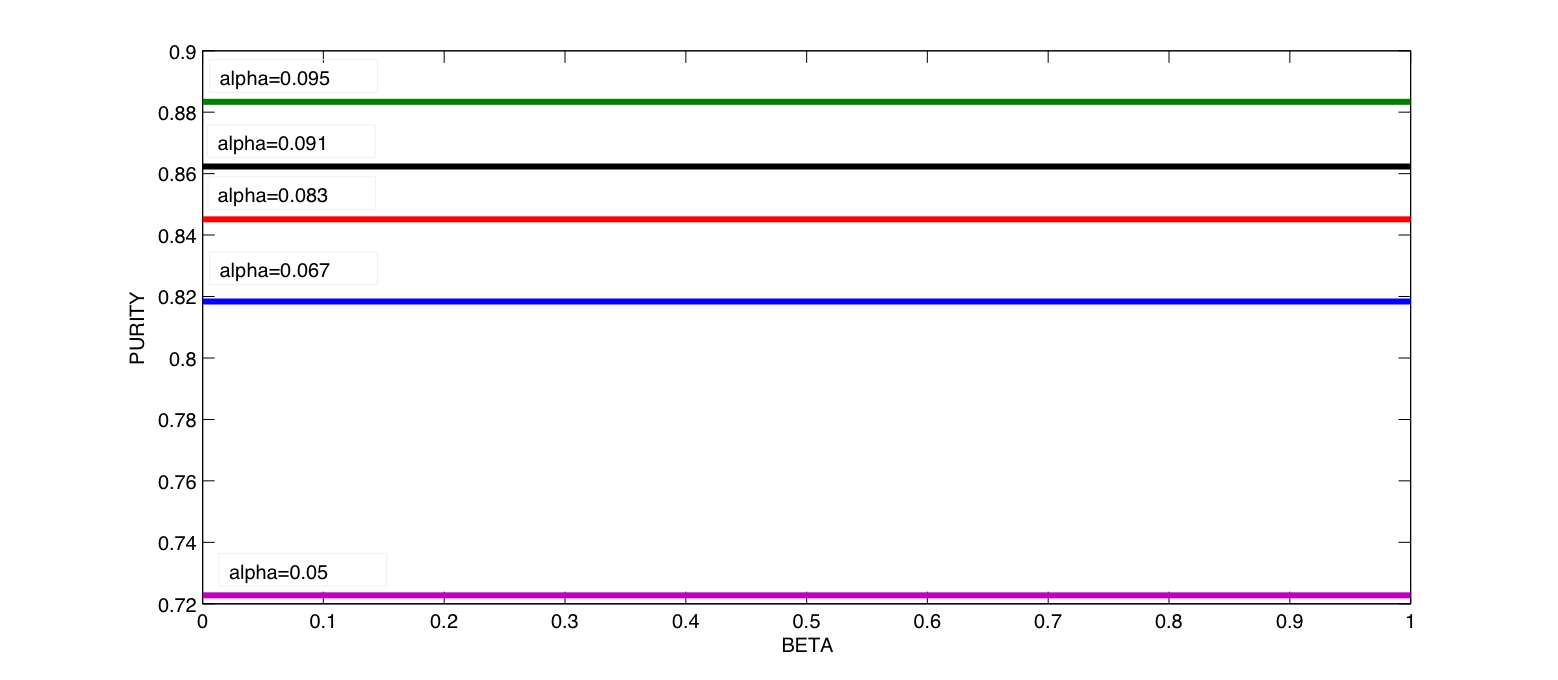}  \\
   (a)\\
   \includegraphics[width=0.55\textwidth]{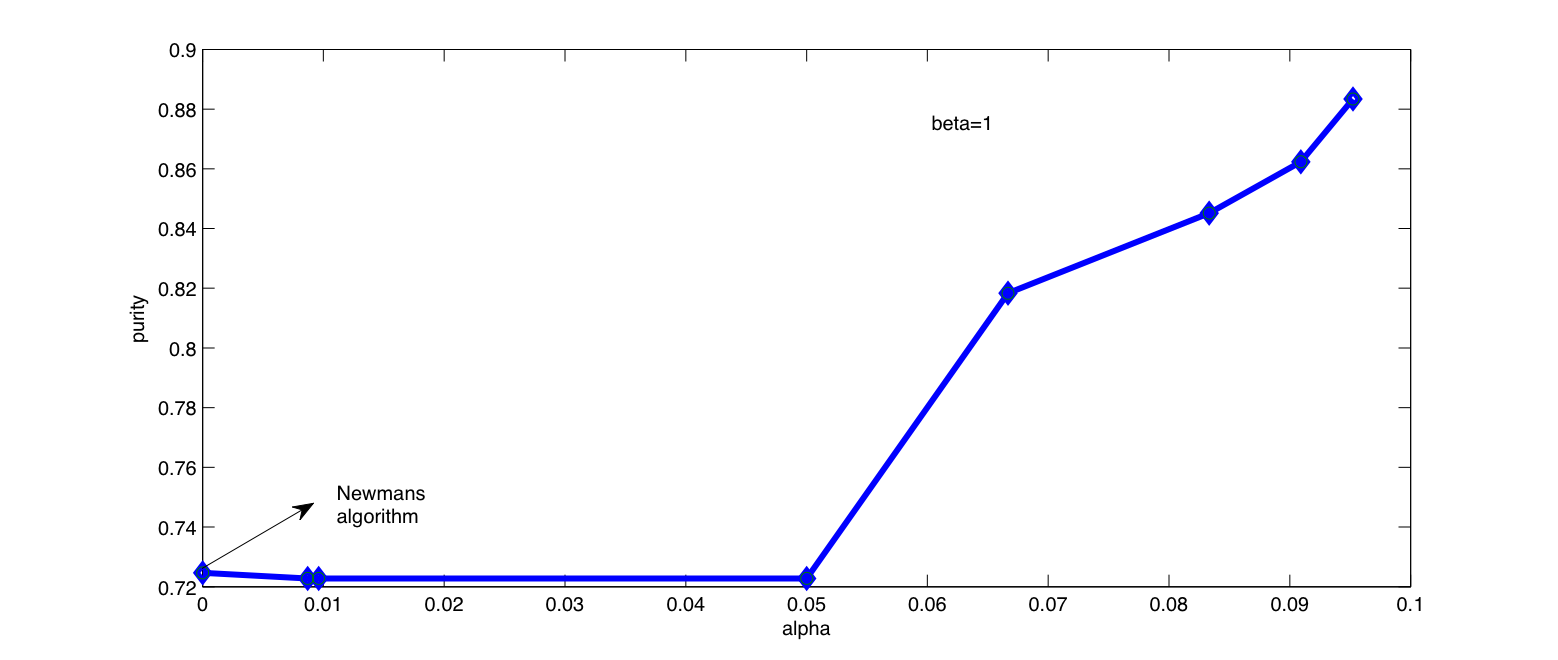} \\
   (b)
 \end{tabular}
 \caption{The graph showing the purity of communities predicted  with different values of $\alpha$ and $\beta$. (a) Case when $\beta$ is varied while keeping $\alpha$ constant. We see that purity is dependent primarily on the value of $\alpha$.
 (b) Case when $\alpha$ is varied and $\beta=1$. We see that as $\alpha$ increases the purity increases reaching to almost $90\%$ near $\alpha=0.1$. This shows that as we increase the attenuated effect of links that are not directly connected to the nodes, the groups become purer. When $\alpha=0$, the method reduces to  eigenvector based modularity maximization method postulated by Newman~\protect\cite{Newman206}.}
 \label{fig:football}
 \end{figure}

We define purity as the total pair-wise similarity between teams that actually belong to the same conference. Thus, the similarity between two teams in a predicted community is 1 if they belong to the same actual conference, and it is 0 it the two teams belong to different conferences. The maximum total similarity would then be obtained if all teams belonging to same conferences end up in the same community. The \emph{purity} of a prediction is then evaluated by the total similarity when teams are grouped in accordance to the communities predicted by the algorithm divided the maximum total similarity. We vary $\beta$ (keeping $\alpha$ constant) and see its change in purity of the predicted communities ~\figref{fig:football}. The graph (\figref{fig:football}(a)) that for a given value of $\alpha$,  purity is constant irrespective of the value of $\beta$, and  hence  purity is dependent primarily on the value of $\alpha$. We next  vary $\alpha$ keeping  $\beta$ constant ($\beta=1$) and compute the corresponding change in purity. \figref{fig:football}(b) shows that community purity increases with the increase of $\alpha$, reaching to almost 90\% near $\alpha=0.1$ (the upper  bound to $\alpha$ is determined by the reciprocal of the largest eigenvalue of the adjacency matrix).  This shows that as we increase the attenuated effect of links not directly connected to the nodes, the groups become purer and it is independent of the attenuated effect of the direct links. When $\alpha=0$ and $\beta=1$,  we get influence dependent only on direct contacts. Hence modularity in this case reduced to one studied by Newman~\cite{Newman206}, and gives around 72\% purity on the football data. The number of groups predicted changes from 8 at $\alpha=0$ to four when $\alpha$ nears 0.1.

\subsection{Political Books}
  \begin{figure}[tbh]
\includegraphics[width=0.55\textwidth]{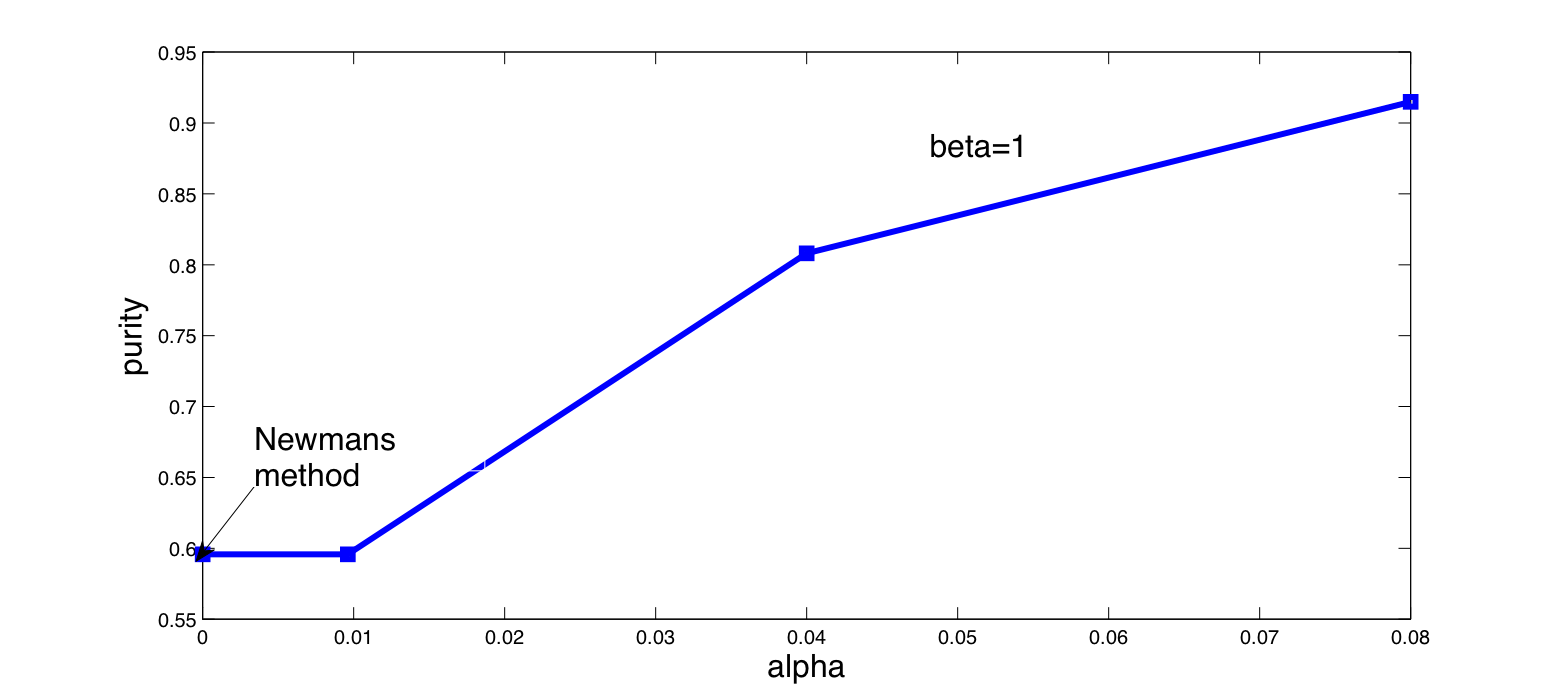}
 \caption{The graph shows the purity of the communities predicted as $\alpha$ is varied ( $\beta$ is kept constant at 1) from $\alpha=0$ to $\alpha=0.08$ (the reciprocal of the largest eigenvalue being taken as the upper bound for the value of $\alpha$.)}
 \label{fig:polbooks}
 \end{figure}

Next we evaluated the approach on the political books data compiled by V.~Krebs.\footnote{\texttt{http://www.orgnet.com/}} In this network the nodes represent books about US politics sold by the online bookseller Amazon. Edges represent frequent co-purchasing of books by the same buyers, as indicated by the ``customers who bought this book also bought these other books'' feature on Amazon. The nodes where given labels \emph{liberal}, \emph{neutral}, or \emph{conservative} by Mark Newman on a reading of the descriptions and reviews of the books posted on Amazon.\footnote{This data is available at \texttt{http://www-personal.umich.edu/$\sim$mejn/netdata/}}. 49 of the books were marked as \emph{conservative}, $43$ books were marked as \emph{liberal} and 13 books were marked as \emph{neutral}. We use our algorithm to find the existing community structure in the network by varying the parameter $\alpha$, as shown in \figref{fig:polbooks}.
We see that as the value of $\alpha$ increases, the number of communities formed decreases (changing from four at $\alpha=0$ to two at $\alpha=0.08$ and keeping $\beta $ constant). Again the reciprocal of the largest eigenvalue being taken as the upper bound for the value of $\alpha$. Also the purity of the communities detected increases from $60\%$ at $\alpha=0$ to as high as $92\%$ at $\alpha=0.08$. Again note that at $\alpha=0$ the method reduces to Newman's modularity maximization method. Another interesting observation is that  when $\alpha$  was taken as $0.08$, leading to the formation of two groups, six of the \emph{neutral} books were  in one group which consisted entirely of \emph{conservative} books (52 books of which 46 were those labeled as  \emph{conservative} and six as \emph{neutral}) and seven were  in the other group (consisting of $53$ books  of  which $43$ were labeled \emph{liberal}, seven were \emph{neutral} and  three were \emph{conservative}). This indicates the possibility that of the 13 books labeled as \emph{neutral}  six  were conservatively inclined and seven were liberally inclined.

\section{Related Research}
\label{sec:related}
Our work is a generalization of the  eigenvector based modularity maximization method proposed by Newman~\cite{Newman206}. Taking $\beta=1$ and $\alpha=0$ reduces the influence matrix to the adjacency matrix, and the modularity that our algorithm maximizes effectively reduces to the modularity defined by Newman~\cite{Newman206}.

The Random Walk models~\cite{tong} and the PageRank algorithm~\cite{PageRank} have been some of the more popular ways of analyzing the relevance of nodes in a network, and may be used for community finding. One way to look at Random Walk models in graph $G(N,E)$ is to start from a vertex $u$ and take random steps along the edges of the graph. The probability of movement from vertex $u$ to $v$ is given by
 \[
T(u,v)= \left \{ \begin {array} {ll}
\frac{1}{d_u} & \mbox{ if  $ \exists$  an edge from vertex $u$ to $v$ in $G$  } ; \\
0 & \mbox{ otherwise}.
\end{array}
\right. \]
where $d_u$ is the degree of vertex $u$. This defines a walk using transition probability matrix $T$.
The second way to look at random walks is to look at probability distribution $\pi_t$ of vertices reached after $t$ steps on traversing the graph $G$. This can be viewed as a probability of  being at a vertex $v \in N$ after time $t$. Let us assume we start from vertex $v_0$, hence the initial probability distribution  of the vertex we are at is
\[
\pi_0(v)= \left \{ \begin {array} {ll}
1 & \mbox{ if  $ v=v_0$  } ; \\
0 & \mbox{ otherwise}.
\end{array}
\right. \]\\
The probability distribution of the vertex that we are at after time $t$ is given by the probability distribution $\pi_t$ and hence
\begin{equation}
\pi_t(v) =\sum_{u \in N} \pi_{t-1}(u)T(u,v)
\end{equation}
This can be represented using $\pi_t = \pi_{t-1}T$;
therefore,
\begin{equation}
\label{ eq:pi2}
\pi_t = \pi_0T^t\,.
\end{equation}

However, for this tool to be useful several factors have to be taken under consideration, including the convergence of the sequence, the stationarity and stability of the distribution, its uniqueness, and so on. If there exists a unique, stable, stationary distribution $\pi$, then this would lead us to
\begin{equation}
\label{eq: pi3}
\pi = \pi{T}
\end{equation}
Computing the eigenvector of the matrix $T$ with eigenvalue 1 gives us the value of $\pi$ which is how  Naive PageRank algorithm evaluates the relevance of the nodes of the network.
Along with the property of the existence of a unique,  stationary, stable distribution, Random Walk with Restart considers an additional probability  that we can return back to our initial state and associates some probability with it.
If we take $\beta$ as the probability to move at random, and $1-\beta$ as the probability of jumping back to its initial state, the Random Walk with restart can be formulated as:
\begin{equation}
\label{eq: pi4}
\pi_i = \beta T \pi_i +(1-\beta) e_i\,,
\end{equation}
where $e_i=[e_{ij}]$ and
\[
e_{ij}= \left \{ \begin {array} {ll}
1 & \mbox{ if  $ j=i$  } ; \\
0 & \mbox{ otherwise}.
\end{array}
\right. \]
Hence vector $\pi_i = [\pi_{ij}] $ gives the relevance score of all nodes $j \in N$ with respect to node $i$.
Similarly, along with the property of the distribution being stationary, PageRank with restarts considers at each time step $t$, probability  $\beta$ to move at random, probability of $1-\beta$ to jump to some specific state, uniformly at random. Hence the transition matrix in this case is modified to $T^{\prime}$ where each element $T_{ij}^{\prime} $ is given by
\begin{equation}
\label{eq: pi5}
T_{ij}^{\prime }= \beta T_{ij} + (1-\beta)/N
\end{equation}
and then as in Equation ~\ref{eq: pi3}  pagerank $\pi$ would then be given by principal eigenvector of this matrix and hence $\pi$ would similarly be
\begin{equation}
\label{eq: pi5a}
\pi= \pi T^{\prime}
\end{equation}

In effect  existence of a unique, stable stationary distribution is the fundamental concept behind most variations of  random walk models and page rank algorithms~\cite{tong,PageRank}. Though widely used especially in the determination of relevance scores, they do have certain limitations.
The non-symmetric nature of a directed graph can lead to problems in the determination of  the unique stable stationary distribution. We have $T=D^{-1}A$ where $D$ is the diagonal matrix of outdegrees. When $G$ is undirected, the adjacency matrix $A$ is symmetric, so the corresponding Laplacian is also symmetric, guaranteeing  favorable properties of the spectrum like the orthonormal basis of real eigenvectors. We can symmetrize $T$ by considering a spectrum of $T+T^T$ or $T \cdot T^T$ (where $T^T$ is the transpose of matrix $T$), but the problem then lies in the graphical interpretation eigenvalues without which these approaches are not really useful. In real life, and in social networks, there do exist directed graphs and as illustrated above it is difficult to apply  the random walk and page rank models on them.


If we think of vertices of the random walk graph as states of a Markov chain, then the property that governs the limiting behavior of $ \pi_0T^t$ is ergocity and we say that the corresponding Markov chain is ergodic if there exists a unique stationary distribution $\pi$ to which $ \pi_0T^t$ converges. The necessary and sufficient conditions of  ergodicity of a Markov chain are irreducibility and aperiodicity. The Random Walk models can be used as a measure of mutual relevance  and PageRank for relevance scores of individuals. However when we consider graphs in real life, especially social networks, these conditions are not necessarily satisfied (e.g., isolated communities).\\

These algorithms are basically concerned with the flow of information on a network. So, if we start from a node with, say a unit of information, which it spreads via the channels it has (outgoing links), the Random Walk model describes the spread of this information in the network  when the information flow attains equilibrium, and further exchange of information among the nodes does not change the distribution of information. When we are thinking of the division of nodes into communities, we are not interested in the amount of information they finally have from each other, but in how this information reaches them, i.e., the channels of the flow of information. The more the channels for information flow a node has, the greater the tendency for the information it sends to reach its recipients. In other words, Random Walk models and PageRank algorithms are concerned with the equilibrium distribution of the flow of information, and we, on the other hand, are interested in the channels of  information flow and their capacity to spread the information.

Mathematically the difference between the two approaches can be stated as follows. Equation\eqref{ eq:pi2} gives us
$\pi_t = \pi_0T^t$.
Let $\pi_{0}(v_i)$ be the vector representing the initial probability distribution  of being there at a particular vertex  when we initially start the random walk from vertex $i$. Obviously in this case we know that we are at $i$ and hence the value of  $\pi_{0}(v_i) $ is given by the unit vector $e_i$ (defined above).
Hence, \\
$[\pi_{0}(v_1), \pi_{0}(v_2), \cdots ,  \pi_{0}(v_N)] =[e_1,e_2,\cdots,e_N]= I$,
where $I$ is the identity matrix.
Hence, if we take \\ $P_t^{'} =[\pi_{t}(v_1), \pi_{t}(v_2), \cdots ,  \pi_{t}(v_N)]$,  where $\pi_{t}(v_i)$ be the vector representing the probability distribution  of reaching the vertices in $t$ steps, when we initially start the random walk from vertex $i$ we have
\begin{eqnarray}
\label {eq:rwr}
P_t^{\prime} & = &I{T^t} \\
& =& {(D^{-1}A)}^{t} 
\end{eqnarray}
The relevance matrix $P^{\prime}$ given by the basic Random Walk model then is Equation\eqref{eq:rwr} at time $t_n$ such that
\begin{equation}
P^{\prime}={P_{t_n}}^{\prime} ={P_{t_{n-1}}}^{\prime} {D^{-1}A}
\end{equation}
On the other hand, we compute the influence matrix $P$  as we have shown above is given by
$P = \beta A {( I-\alpha A)}^{-1}$.

We can compute the  influence score of the nodes relative the network  using  the influence matrix as done by Katz ~\cite{Katz}. Taking $p_{ij}$ as the influence scores of the nodes with respect to each other, i.e., $P_{ij} =p_{ij}$, we have $p_i = \sum_{j} p_{ij}$. Hence, the column vector $p$ whose elements are $p_i$ gives the influence score of the nodes relative to the network.

Recently researchers have applied probabilistic models, such as mixture models, to the community discovery task. The advantage of these models is that can probabilistically assign a node to more than one community, because, as it has been observed ``objects can exhibit several distinct identities in their relational patterns''~\cite{Airoldi,tina}. This indeed maybe true, but whether the nodes in the network is to be divided into distinct communities or probabilities with which each node belongs to  community is to be discovered, really depends on the specific application.

\section{Conclusion and Future Work}
\label{sec:conclusion}
We have proposed a new definition of a community in terms of the influence that nodes have on each other. We gave a mathematical formulation of influence in terms of the number of paths of any length that link two nodes, and redefined modularity in terms of the influence metric. We use the new definition of modularity to partition a network into communities. We applied this framework to networks well-studied in literature and found that it produces results at least as good as the edge-based modularity approach.

Although the formulation developed in this paper applies equally well to directed graphs, we have only implemented the algorithm on undirected ones. Hence future work includes implementation of the of the algorithm on directed graphs that are common on social networking sites, as well applying it to bigger networks.

Leskovec et al.~\cite{Leskovec08www} state that they ``observe tight but almost trivial communities at very small scales, the best possible communities gradually `blend in'  with rest of the network  and thus become less `community-like'.'' However the hypothesis that they employ to detect communities  is that communities have ``more and/or better-connected `internal edges' connecting members of the set than `cut edges' connecting to the rest of the world.'' Hence, like most graph partitioning and modularity based approaches to community detection, their process depends on the local property of connectivity of nodes to neighbors via edges and is not dependent on the structure of the network on the whole. Besides, it also  does not take into account the heterogeneity of node types, that is `who' are the nodes that a node is connected to and how influential these nodes are. Therefore, we argue that a global property, such as the measure of influence, is a better approach to community detection. It remains to be seen whether communities will similarly `blend in' with the larger network if one uses the influence metric to discriminate them.

\subsection*{Acknowledgements}
This research is based on work supported
in part by the National Science Foundation under Award Nos.
IIS-0535182, BCS-0527725 and IIS-0413321.

\bibliographystyle{plain}
\bibliography{references}

\appendix{}
Below we summarize the application of the leading eigenvector method of Newman~\cite{Newman206} to influence-based modularity. If we consider the division of the network into two communities, then we could write $Q$ as :
 \begin{equation}
\label{eq: mod16}
Q=\sum_{ij} {(P_{ij} - \frac { {W_{i}^{out}} {W_{j}^{in}}}{W})(\frac{s_{i}s_{j}+1}{2})}= \frac {1}{2} \big(s^{T}Cs +\sum_{ij}C_{ij}\big)
\end{equation}
where \\
\[
s_i = \left \{ \begin {array} {ll}
1 & \mbox{ if vertex $i$  $\in$ group 1} ; \\
-1 & \mbox{ if vertex $i$  $\in$ group 2}.
\end{array}
\right. \]
and $s$ is a vector whose elements are $s_i$ and matrix $C$ comprises of elements $C_{ij}$ such that
$C_{ij} = P_{ij} - \frac { {W_{i}^{out}} {W_{j}^{in}}}{W}$.
We symmetrize matrix $C$ to get matrix $B = C+C^{T}$. $B$ is now called the \emph {modularity matrix}, and we approximate modularity as
\begin{equation}
\label{eq: mod17}
Q=\frac{1}{2} \big( s^{T}Bs +\sum_{ij} B_{ij} \big)
\end{equation}

Hence if we want to divide the network in such a way that there is more than expected influence within the communities, we would have to maximize the change in modularity due to  subdivision. We note that before the initial division, i.e., taking the entire network,  since  all the elements belong to the same community or group the modularity is $Q=\sum_{ij} B_{ij}$.
Therefore, additional contribution $\Delta Q$ to modularity upon dividing subgroup  $g$ is:
\begin{eqnarray}
\label{eq: mod19}
\Delta Q &=& [( \sum_{i,j \in g} {B_ij (\frac{s_{i}s_{j} + 1}{2})})-( \sum_{i,j \in g} {B_ij })] \\
\Delta Q &=& \frac{1}{2}[( \sum_{i,j \in g} {B_ij (s_{i}s_{j})-( \sum_{i,j \in g} {B_ij })}] \\
\Delta Q &=&  \frac{1}{2} \sum_{i,j \in g}[ B_{ij}- \delta_{ij} \sum_{k \in g}B_{i,k} ]s_{i}s_{j}\\
\Delta Q &=&  \frac{1}{2} s^{T}B^{(g)}s
\end{eqnarray}
where
$B_{ij}^{(g)}= B_{ij}- \delta_{ij} \sum_{k \in g} B_{ik}$
and $ g $ is the entire network  for the first division of the directed graph into two communities $C_1$  and $C_2$. We can iteratively subdivide the resulting communities $C_1$ and $C_2$. $\Delta Q$  reflects the additional contribution to modularity of the entire network as the result of  these subdivisions. If no  further division increases modularity,  we stop the process. The communities thus found are the optimal, or natural, communities within the network.

Next we show that maximizing the modularity can be  approximated using eigenvalue decomposition.
We  can write $s$ as a linear combination of the normalized eigenvectors $u_i$  of  $B^{(g)}$.
Hence
\begin{equation}
\label{eq: mod20}
s=\sum_{i} a_{i} u_{i}
\end{equation}
Hence $a_{i} = u_{i}^{T}.s$ \\
therefore
\begin{eqnarray}
\label{eq: mod21}
\Delta Q= \frac{1}{2} s^{T}B^{(g)}s\\
\Delta Q= \frac{1}{2} {(\sum_{i} a_{i} u_{i})}^{T}B^{(g)}(\sum_{i} a_{i} u_{i})\\
\Delta Q= \frac{1}{2}\sum_{i} a_{i}^{2} u_{i}^{T}B^{(g)}u_{i}\\
\Delta Q= \frac{1}{2}\sum_{i} a_{i}^{2} \lambda_{i}\\
\Delta Q= \frac{1}{2}\sum_{i}(u_{i}^{T}.s )^{2} \lambda_{i}
\end{eqnarray}
where $ \lambda_{i}$ is the eigenvalue of $B^{(g)}$ corresponding to  eigen vectors $u_i$.The eigenvalues (and their corresponding eigenvectors) are labeled in decreasing order of their magnitude  \ie
$\lambda_{1} \geq \lambda_{2} \geq \lambda_{3} \geq \lambda_{4}  \geq \cdots $

Since we wish to maximize $\Delta Q$  hence we would like to choose the value of $s$ such that maximum weight is concentrated on the largest eigen values. The optimized solution would then be to choose $s$ proportional to $u_1$. However the constraint in choosing s in this manner is that s has an additional constraint that it can only  be eiher 1 or -1. The approximation then used is similar to the one used  spectral partitioning where all nodes whose corresponding elements in $u_1$  are positive put in one group and the rest in the other group.

\end{document}